# Which Discoveries Are Paradigm Shifting?

Sajad **Ashouri**[1,2] (sajad.ashouri@vtt.fi), Arash **Hajikhani**[1] (arash.hajikhani@vtt.fi), Ari **Hyytinen**[3] * (ari.hyytinen@hanken.fi), Petri **Rouvinen**[4] (petri.rouvinen@etla.fi), Arho **Suominen**[1,2] (arho.suominen@vtt.fi)

1: VTT Technical Research Centre of Finland (Tekniikantie 21, 02150 Espoo)
2: Industrial Engineering & Management, Tampere University, Finland (Kalevantie 4, 33100 Tampere)
3: Hanken School of Economics, Finland (Arkadiankatu 22, 00100 Helsinki), and
   Helsinki GSE, Finland (Arkadiankatu 7, 00100 Helsinki)
4: ETLA Economic Research, Finland (Arkadiankatu 23, 00100 Helsinki)

* **Corresponding author**: Ari Hyytinen.

**Abstract**: To better align theories of paradigm shifting discoveries and empirics identifying them, we propose a novel measure that incorporates a discovery's impact, novelty, and tendency to break with the past into a single, coherent measure. Calibration using the National Inventor Hall of Fame data reveals that impact, novelty, and disruptiveness are strict complements meaning, for example, that greater impact cannot substitute for moderate novelty. We illustrate the workings of the measure using data on USPTO patents from 1982 to 2015.

**Acknowledgements**: We thank Giacomo Calzolari, Vincenzo Denicolò, Russell Funk, Tuomas Takalo, Otto Toivanen, Roope Uusitalo and seminar participants at MaCCI Annual Conference 2024 and the Annual Meeting of the Finnish Economic Association 2024 for useful discussions. On his part, Hyytinen thanks Yrjö Jahnsson Foundation, FORTE (grant no. 2021-01552_3), Saara and Björn Wahlroos' foundation and dr. h.c. Marcus Wallenberg's Foundation for Research in Business Administration for financial support. Key parts of this research were completed while Hyytinen was visiting University of Bologna, Italy.

Published version: Sajad Ashouri, Arash Hajikhani, Ari Hyytinen, Petri Rouvinen, Arho Suominen, Which discoveries are paradigm shifting?, Industrial and Corporate Change, 2026;, dtag002, https://doi.org/10.1093/icc/dtag002

# 1. INTRODUCTION

Researchers stand on the shoulders of giants, but what determines on whose shoulders we stand in a particular endeavor? When research advances incrementally, it can be conceptualized as a continuous process that recombines existing knowledge (Bresnahan, 2012; Fleming & Sorenson, 2001; Weitzman, 1998) and that enables research trajectories that are cumulative, integrative, and path dependent (Augsdorfer, 2005; Patel & Pavitt, 1997). This progress corresponds to 'normal science' (Kuhn, 1962; see also Leahey et al., 2023), leading often to piecemeal scientific and technological progress. Yet, discoveries can also be disruptive (e.g., Christensen, 1997; Christensen et al., 2002; Felin & Kauffman, 2023; Hopster, 2021; Schmidt & Druehl, 2008), leading to discontinuities, to restructuring of existing organizations and markets (Adner, 2002; Arthur, 2007, 2009; Christensen et al., 2018; Christensen et al., 2015; O'Connor & Veryzer, 2001), and to abrupt changes in the evolution of sociotechnical systems (e.g., Kivimaa et al., 2021; Wilson & Tyfield, 2018). Such paradigm shifting discoveries are rare (Dosi, 1982; Kuhn, 1962; Wuestman et al., 2020), but when they happen, they tend to break with the past, thus altering the shoulders on which subsequent research is built. The contribution of this study is that we propose a novel measure that simultaneously captures three salient dimensions of such paradigm shift discoveries, i.e., their tendency (i) to have a significant impact on subsequent inquiry, (ii) to be novel in how they use prior knowledge, and (iii) to break with the past (i.e., to render some previous discoveries obsolete).

Several scholars have contributed to the recent debates on the nature of scientific and technological advance and on the tendency of discoveries to break with the past.[1] In their seminal contribution, Funk and Owen-Smith (2017) introduced the Consolidation-Disruption index (the CD index) to capture discontinuities in the trajectory of technological development. Their key insight is that when a disruptive technology provides a platform for future discoveries, it changes how future discoveries cite previous ones. Wu et al. (2019) apply this measure to patent and publication data, arguing that the measure identifies successfully both disruptive inventions and scholarly writings. Recently, Park et al. (2023) applied the CD index to explore whether research articles and patents have become less likely to break with the past (see also Macher et al., 2024). In addition, there are several studies that have used the CD index to examine the tendency of discoveries to break with the past using field or sector specific data; for a review, see Leibel and Bornmann (2023). Yet, despite its importance for the theories of scientific and technological change and its practical relevance for, e.g., managerial decisions and corporate strategies, measuring the nature of this progress, especially its disruptiveness, is still subject to considerable debate (Funk & Owen-Smith, 2017; Leahey et al., 2023) and lacks consensus (Bornmann et al., 2020; Leibel & Bornmann, 2023; Nagy et al., 2016).[2] Leibel and Bornmann (2023) discuss in detail the ongoing controversies in how to measure disruptiveness using citations data. An example of an alternative approach to the citation-based measures is Kelly et al. (2021), who utilize textual analysis of patent documents to deduce indicators of patent quality and innovation.

---

[1] The literature refers to exceptional and unusually significant innovations using various labels, such as "radical" (Capponi et al., 2022; Coccia, 2017; Dosi, 1982), and "breakthrough" (Barnholt, 1997; Mascitelli, 2000).

Inspired by the seminal contribution of Funk and Owen-Smith (2017) and by the recent work of Leahey et al. (2023), our aim is to better align theoretical work on abrupt scientific and technological advancements, manifesting themselves as paradigm shifting discoveries (Kuhn, 1962; Wuestman et al., 2020), and empirical work aiming to identify and to quantify them. Whereas some fundamental paradigm shifting discoveries may change how human beings understand the world (e.g., Copernican revolution that led to heliocentrism), we seek to quantify paradigm shifts in their narrower meaning, i.e., when they concern a broadly defined scientific discipline or technological field. While narrower, such paradigm shifts create new types of external economies and spillovers, reshaping thus both firms' internal R&D projects and collaborative and competitive interaction between firms within broader industrial contexts, as discussed by Sturgeon (2002). Building on the views put forward by, e.g., Leibel and Bornmann (2023) and Leahey et al. (2023), we argue that existing (mostly unidimensional) measures cannot successfully capture the tendency of such paradigm shifting discoveries (i) to be highly cited, (ii) to be novel, and (iii) to render some of the previous discoveries obsolete. To wit, citation counts mirror the strength of a discovery's impact, but not its nature. On the other hand, the widely-used CD index of

---

[2] Echoing this, Park et al. (2023, p. 143) note that the CD index "will benefit from future work on its behaviour and properties".

Funk and Owen-Smith (2017) and its subsequent modifications (e.g., Bornmann et al., 2020; Leydesdorff et al., 2021; Wang et al., 2023; Wei et al., 2023) aim at measuring whether a discovery alters how preceding findings and technologies are cited. These measures do not capture a discovery's impact, since a patent or a research paper can obtain a high CD index value and may still receive only a small number of forward citations. This suggests that it has barely been noticed by the relevant community and is unlikely to redirect the course of scientific and technological advance (Bu et al., 2021; Funk & Owen-Smith, 2017; Leibel & Bornmann, 2023; Wei et al., 2023). Finally, theories of scientific change suggest that a primary reason why discoveries tend to break with the past is their greater novelty, since such findings are associated with uniquely creative ideas and tend to draw attention away from past efforts (Foster et al., 2015; Leahey et al., 2023; Lee & Min, 2015; Leibel & Bornmann, 2023; Sheng et al., 2023; Uzzi et al., 2013).

This paper builds on the insight that to empirically single out paradigm shifting discoveries, we need a single, coherent measure that incorporates their three identifying dimensions, i.e., their impact, novelty, and tendency to break with the past (Kuhn, 1962; Wuestman et al., 2020). We propose that, when calculated over these three identifying dimensions, the well-known mathematical concept of generalized mean (e.g., Bullen, 2003) leads into such a measure. While the generalized mean has been used in applied work earlier (for applications and outcomes, see, e.g., Ahmadpoor & Jones, 2019; Haley, 2019; Perry & Reny, 2016), our application of it is novel. A further novelty is that we normalize the dimensions over which the generalized mean is calculated using the percentages of their respective cumulative distributions (Bornmann et al., 2020; Leydesdorff et al., 2021; Park et al., 2023). Importantly, this normalization means that the complete distribution of each variable is considered (Bornmann & Williams, 2020), enabling researchers to *rank* patents or publications relative to their comparators in terms of their impact, novelty, and tendency to break with the past.

As we demonstrate, the proposed measure has several strengths. First, it allows ranking discoveries across the three identifying dimensions and then compresses them into a single, easy-to-understand, quantitative measure that varies on a well-defined interval from zero to one. Similarly to the graph theory-based approach of Gebhart and Funk (2023), this nested approach considers simultaneously the number of citations, their novelty, and their interconnectedness. This is a desirable property, because empirically the identifying dimensions are only moderately (if at all) correlated and thus contain independent information. Second, because of its flexible parametrization, the measure does not restrict a priori whether discoveries' impact, novelty, and tendency to break with the past are complements or substitutes. To infer what the appropriate parametrization is, one can use either theoretical arguments (i.e., axiomatically, as in Perry & Reny, 2016) or empirical analysis (as in Ahmadpoor & Jones, 2019), depending on the goal of the study. In our application we opt for using both. Specifically, we calibrate the proposed measure using data on USPTO patents and significant discoveries made by the inventors listed in the National Inventor Hall of Fame (NIHF). The patents recognized by NIHF can be interpreted to represent paradigm shifting discoveries because they have laid an acknowledged foundation for new discoveries or revolutionized existing ones. Our calibration reveals that the three dimensions (impact, novelty, disruptiveness) are strict complements, which suggests, e.g., that a greater tendency of a discovery to break with the past cannot compensate for its weak impact in determining whether it is paradigm shifting. The calibrated measure can be used to predict paradigm shifting discoveries in unseen data (not used in calibration). Finally, the measure can be further extended to, e.g., to include yet other dimensions of paradigm shifting discoveries, implying that it can be tailored and re-calibrated to new research contexts in a transparent way. Because of these properties, we consider the proposed measure flexible and call it $D_{FLEX}$ for brevity.

There are two further motivations to use $D_{FLEX}$ when identifying unusually significant discoveries. On one hand, applications of the existing disruption measures (e.g., the CD index and its modifications) seem to implicitly assume that they also capture something greater than just discoveries' tendency to break with the past in a technical sense, such as their capacity to revolutionize certain fields of science or technology. The literature seems to contain only a limited number of suggestions of how such capacity could be captured. One suggested solution is the citation-weighted CD index (the mCD index) of Funk and Owen-Smith (2017), which is obtained by multiplying a discovery's CD index by its citation count. However, this measure is sensitive to variation in the citation counts due to the different scale of the CD index and the citation count (Leibel & Bornmann, 2023; Wei et al., 2023).³ The proposed $D_{FLEX}$ neither suffers from the scale issue nor requires using separate cut-offs.⁴ Moreover, contrary to the existing measures, our application of $D_{FLEX}$ would detect that the USPTO patent on polymerase chain reaction, protecting a founding invention in modern biotechnology, is unique in terms of impact and disruptiveness but is not as novel. On the other hand, $D_{FLEX}$ improves upon exiting measures, because it naturally draws attention to how rare paradigm shifting discoveries are, i.e., to them being fundamentally a *tail* phenomenon (e.g., Fleming, 2007; Kaplan & Vakili, 2015; Kelly et al., 2021; Kuhn, 1962). Specifically, paradigm shifting inventions should be sufficiently unique when compared to others in terms of their impact, novelty, and disruptiveness and also less commonly observed than their slightly less unique counterparts. This property does not apply to the CD index and its recent modifications, because its empirical distribution tends to have a concentration at the right tail (see Bornmann et al., 2020; Holst et al., 2024). Possible measurement issues aside (Holst et al., 2024), it is not theoretically plausible that there would be many times more truly disruptive research findings (obtaining the maximum value of the measure) than findings that are only *slightly less* so (obtaining a value just below the maximum). Being multidimensional, the proposed $D_{FLEX}$ allows for a more fine-tuned analysis of the discoveries that otherwise would tend to cluster implausibly based on the existing measures.

We apply the $D_{FLEX}$ measure to the USPTO patent data from 1982 to 2015 and to discoveries made by the inventors listed in the NIHF. Our estimations show that when considering which discoveries are likely to be paradigm shifting, their impact, novelty, and tendency to break with the past are strictly complementary. This property means, for example, that a discovery's great impact cannot compensate for its moderate novelty or disruptiveness. This illustrates concretely the gain of putting the three measures together as opposed to using them in isolation; we find in our empirical application that it is the weakest of the three dimensions that determines the tendency of a discovery to be paradigm shifting. In additional analyses, the measure predicts significant patents, as captured by their long-term citation counts. $D_{FLEX}$ also provides further insights of a selection of patented inventions that have been considered significant and disruptive in previous literature. Finally, we illustrate how $D_{FLEX}$ can be used by applying it to USPTO data from 1982 to 2015.

---

³ The difference in the scale of the two components of the citation-weighted CD index means that highly cited discoveries with CD index values relatively close to zero may, counterintuitively, end up appearing extremely disruptive (or consolidating) in light of the mCD index. As a remedy, Wei et al. (2023) propose keeping the CD index and citation counts separate and using multiple cutoff values to determine which discoveries are consistent with revolutionary science.

⁴ The proposed normalization suing the percentages of the respective cumulative distributions removes this problem. In fact, we can show that a rescaled version of the mCD index that can be obtained as a special limit case of $D_{FLEX}$. This rescaled version can be applied to data without using multiple hard-to-define cut-off values, such as those used in Wei et al. (2023).

The rest of this article is as follows: In Section 2, we outline a conceptual framework for our analysis. In Section 3, we review earlier literature on the measurement of discoveries' disruptiveness and then introduce a new, multidimensional measure for identifying paradigm shifting discoveries. In section 4, we turn to empirical analysis, using the USPTO patent data from 1982 to 2015. Therein, we illustrate some of the challenges with existing disruptiveness measures, how the new measure can be parametrized, and provide examples of how it works. Section 5 discusses, and Section 6 concludes.

## 2. CONCEPTUAL FRAMEWORK

### 2.1. Scientific revolutions and paradigm shifting discoveries

As seminal works by Kuhn (1962), Laudan (1978), and Whitley (2000) suggest, revolutionary or paradigm shifting discoveries tend (i) to have a major impact on subsequent inquiry, (ii) to be novel in how they use and combine prior knowledge, and (iii) to render previous knowledge obsolete and to break with past trajectories. Such discoveries change the underlying assumptions, concepts, explanations, and interpretations within a scientific discipline, leading to a new and more comprehensive understanding of the central phenomena that the discipline purports to explain. By doing so these discoveries redirect and interrupt the normal course of scientific and draw attention away from the past research while influencing subsequent research efforts (Funk & Owen-Smith, 2017; Leahey et al., 2023; Leibel & Bornmann, 2023; Sheng et al., 2023; Wei et al., 2023). These views suggest that, rather than focusing on disruptiveness alone, scientific revolutions need to be scrutinized as paradigm shifts, i.e., as something that serve as a basis for (ample) subsequent efforts, involve a sufficient degree of originality and creativity (novelty), and bring about discontinuity.

At the time of Kuhn's (1962) contribution and also during subsequent decades, science and technology were seen as much less overlapping than what they are considered today. Dosi (1982) broadened Kuhn's (1962) notion of paradigm shifts in science to technological change, noting that "*Technological*

*paradigms and trajectories are in some respects metaphors of the interplay between continuity and ruptures in the process of incorporation of knowledge and technology into industrial growth…*" (p. 161).[5] While Dosi seems to emphasize a bit more the continuous nature of technological advance, his thinking is in line with Kuhn's, to whom Dosi draws direct parallels in his characterization of technological paradigms and trajectories. While we do not yet "*fully understand what technological paradigms are*" (von Tunzelmann et al., 2008, p. 481; see also Nelson, 2008), broadly speaking they are overarching frameworks guiding research and development efforts and technological inquiries. Discontinuities and shifts in these frameworks arguably originate from discoveries that tend to be technologically unique and significant in how they change those efforts and inquiries. Yaqub (2017) explores how technological paradigms shape technological trajectories and the impact of significant discoveries on the evolution of industries, using the examples of vaccines and HIV vaccines. This work highlights the importance of understanding how paradigm shifting discoveries can alter the course of technological progress within specific industrial contexts.

Whereas it may be tempting to equate paradigm shifting discoveries with fundamental changes in how human beings view and understand the world, such as the Copernican revolution that resulted in a shift towards heliocentrism or discoveries of Newtonian mechanics and quantum mechanics, we seek to quantify discoveries that are paradigm shifting in a somewhat narrower sense. We focus on paradigm shifts induced by single inventions or discoveries, as opposed to a cluster of related discoveries, *within* a given broadly defined scientific discipline or technological field. Such field-specific discoveries challenge and potentially replace the prevailing scientific consensus on a major topic, forcing researchers and developers of the field to re-evaluate existing theories, develop new explanations and rethink the types of questions they need to study. These kinds of discoveries do not just add new information but potentially disrupt existing knowledge by offering alternative explanations that require researchers to re-evaluate the prevailing consensus in the field in question and perhaps beyond it. Sturgeon (2002) highlights how measuring the impact and novelty of discoveries within their broader industrial context has become increasingly important, reflecting the growing role of inter-firm interactions and external economies in driving innovation.

---

[5] The lasting legacy of Dosi's (1982) seminal contribution is discussed at length by von Tunzelmann et al. (2008) in an introduction to a special issue and in its articles (in particular, see Nelson, 2008).

We argue, moreover, that be they scientific or technological, paradigm shifting discoveries are fundamentally *a tail phenomenon* (Kelly et al., 2021; Kuhn, 1962; Wuestman et al., 2020). This suggests that we should consider explicitly how a discovery, such as a patented invention, ranks relative to others; paradigm shifting discoveries are likely to be unique when compared to current and former findings within broadly defined scientific disciplines and technological fields (Dahlin & Behrens, 2005). Because a paradigm shifting discovery is likely to be dissimilar in the considered dimensions when compared to others, observing them ought to be a very infrequent (low frequency) occurrence. Therefore, irrespective of the specifics of the measure, it is natural to focus on the tail(s) of its distribution (Kelly et al., 2021) when considering which discoveries are paradigm shifting.

## 2.2. Multidimensionality of paradigm shifting discoveries

Discoveries and innovations are heterogenous and vary in their degree of complexity and structure (Dahlin & Behrens, 2005; Fleming, 2007; Higham et al., 2021; Scherer & Harhoff, 2000), suggesting that the capacity of a discovery to be paradigm shifting is multidimensional. Leahey et al. (2023) similarly argue that unidimensional measures are unable to holistically capture factors associated with novelty and disruption (see also Bu et al., 2021; Chen et al., 2021). A key argument of this paper is that we ought to focus on the *combined* ability of impact, novelty, and disruptiveness to encapsulate empirically paradigm shifting discoveries.

Whereas it has been acknowledged that impact, novelty, and disruptiveness can be interviewed and partly overlapping (Leahey et al., 2023; Strumsky & Lobo, 2015; Verhoeven et al., 2016), they are distinct concepts. Reflecting this, their meaning, measurement, utility, and interpretations have been discussed in a somewhat disconnected fashion in the literature. For simplicity, we also treat them separately and address each concept in turn:

**Impact**: Impact refers to the scale of influence a research outcome has on subsequent inquiry. Measuring such an impact is a multifaceted endeavor. One methodology for gauging impact is to scrutinize citations that research articles or patents receive. In our context, the number of forward citations that a patent obtains is the most apparent ex-post way of identifying a discovery's impact. Patents with many forward citations are, by a standard revealed preference argument, important for future inventions: the higher the number of forward citations, the more important the patented invention is for subsequent inquiry (Hall et al., 2001; Harhoff et al., 1999). Moreover, the sooner a patent accumulates a significant number of citations, the more impactful it is. Forward citations also reflect inventions' private and social value (Bloom et al., 2013; Harhoff et al., 2003; Kline et al., 2019; Kogan et al., 2017). What citations cannot fully capture is that an impactful outcome may also have implications beyond the sphere of research. It may, e.g., shift business practices, change competition in the marketplace, or lead to regulatory changes (Bornmann, 2012, 2016; Brueton et al., 2014; Searles et al., 2016).[6]

**Novelty**: Novelty refers to the quality of being original or unprecedented. It denotes previously unexplored or undocumented ideas or new combinations of old ideas. Novelty is a necessary condition for humankind's collective advance of knowledge. In practice, measuring novelty involves comparing an idea to the body of existing knowledge. What differs from what is already established, is considered novel. In line with this view, an unprecedented combination of technological areas spanned by a patent has been associated with novelty (Strumsky & Lobo, 2015; Verhoeven et al., 2016). This view builds on the premise that an invention introducing new types of co-occurrences of distinct areas tends to be a novel recombination of existing knowledge (e.g., Weitzman, 1998).

Strumsky and Lobo (2015) argue that a patent's novelty can be inferred from the technological classification of patents. In this view, the novelty of a patent has three levels – origination, combination, and refinement. Origination implies the highest degree of novelty, since it means that an innovation is so novel that a new technological area (appearing in the patent) needs to be introduced for it. The next level, combination, occurs when a novel pairwise combination of technological areas is observed in the patent. Finally, if a patent does not include any novel combinations of technological areas, it is incremental innovation; such patents are refinements. Echoing this line of thinking, Verhoeven et al. (2016)

---

[6] Kelly et al. (2021) provide a specific operational definition of impact in the domain of patenting. They ascertain impact by analyzing a patent's sway over forthcoming inventions. This is quantified by examining a patent's "forward similarity" to all patents lodged within the ensuing five-year timeframe. A heightened forward similarity denotes a pronounced effect of the patent on subsequent inquiry. Park et al. (2023) employ a similar metric. In their approach, impact is gauged by examining how closely a given patent aligns with subsequent patents registered within a five-year period. A surging forward similarity suggests that the patent in question was not just a product of its time but also a catalyst for shaping future inventions.

consider the degree to which a patent includes new combinations of fields in its non-patent references. The more unprecedented combinations there are, the more novel the patent.

Patent documents' textual data is yet another source for inferring novelty (Arts et al., 2021; Bloom et al., 2021; Kaplan & Vakili, 2015; Kelly et al., 2021). For example, Arts et al. (2021) examine patents' title, abstract, and claims to identify novelty, as indicated by new technical keywords, bigrams, and trigrams. A large number of new keyword combinations relative to the other patents issued in the same technology class is indicative of the focal patent being novel. This kind of keyword analysis allows capturing novelty in a more flexible fashion than an analysis of predefined and administratively determined technological fields or classes. Our proposed measure can accommodate both ways of measuring novelty, but Kuhn's (1962) notion that ideas are embedded in words – i.e., in the text of scholarly literature or a patent – suggests using a text-based approach.

**Disruptiveness:** Disruptiveness concerns the capacity of innovations or ideas to alter the trajectory of knowledge accumulation and to break with the past (e.g., Funk & Owen-Smith, 2017; Park et al., 2023; Wu et al., 2019). Rather than being an addition to the existing corpus of knowledge, a disruptive discovery presents a divergence, often rendering prior knowledge redundant and carving out novel paths for subsequent inquiries. This concept is encapsulated by Rosenkopf and Nerkar (2001), who emphasize that disruptiveness represents the capacity of an idea to both challenge the status quo and to set the stage for future explorations. Such a break from past knowledge induces a shift away from what was previously known. Funk and Owen-Smith (2017) delve into this idea and suggest that truly disruptive discoveries not only lay the foundation for future discoveries but also lessen the dependence on earlier ones. Building upon these conceptual points of departure, Christensen et al. (2018), Cozzolino et al. (2018), Hopp et al. (2018), Markides (2006), and Pinkse et al. (2014) delve into the details of disruption. Leibel and Bornmann (2023) review the concept and its measurement.

**Complementarity vs. substitutability:** As we have argued above, a paradigm shifting discovery is likely to have a major impact on subsequent inquiry, to be novel, and to break with past trajectories (Dosi, 1982; Funk & Owen-Smith, 2017; Kuhn, 1962; Laudan, 1978; Whitley, 2000). Whereas the theoretical considerations suggests that a focus on the combined effect of these three dimensions is warranted, theoretical reasoning alone cannot determine whether they should be treated as substitutes or complements. For example, when determining whether a focal discovery is paradigm shifting, it is possible that its lower novelty could be compensated for by the discovery's greater impact or disruptiveness. In cases like this, the three dimensions could be interpreted to be substitutes. However, they can also be complements. In this case, whether a discovery is paradigm shifting depends more on its weakest dimension. For example, a lower degree of novelty of the focal discovery can become a constraint when it cannot be compensated for by the other, possibly better dimensions, such as greater impact. As we will argue in greater detail below, since Kuhn (1962), the available literature has mostly treated the different dimensions as complements when discussing revolutionary science and significant innovations (see, e.g., Bornmann & Williams, 2020; Funk & Owen-Smith, 2017; Leahey et al., 2023; Leibel & Bornmann, 2023; Wei et al., 2023). In our empirical analysis, we use the discoveries of the inventors listed in the NIHF to determine whether complementarity or substitutability is supported by the data.

## 3. MEASURING PARADIGM SHIFTING DISCOVERIES

In this section, we turn to measurement. We start with a concise synopsis of the existing disruption measures; Leibel and Bornmann (2023) offer a more comprehensive overview of the existing disruption measures, including their conceptual background and strengths and weaknesses. Despite insightful recent advances in the measurement of discoveries' disruptiveness, there seems to have been only a few attempts to simultaneously quantify the impact, novelty, and/or disruptiveness of a discovery. We therefore proceed to consider how paradigm shifting discoveries can be quantified based on these three dimensions using a single, coherent measure.

### 3.1. Existing disruptiveness measures

**Unidimensional measures:** A discovery can be consolidating or disruptive, depending on how subsequent discoveries (e.g., patents or publications) cite it and its prior art. Based on this insight, Funk and Owen-Smith (2017) propose identifying disruptive discoveries by using their CD index that measures the extent to which the focal discovery changes subsequent citations. Specifically, if a discovery breaks with the past and renders prior art obsolete, it is disruptive; see also Wu et al. (2019) and Park et al. (2023) who – among others – make use of this interpretation.

To introduce the CD index of Funk and Owen-Smith (2017) more formally, we turn to Figure 1, concretely showing how subsequent patents cite the focal patent and its prior art. In this figure, $N_i$ is the number of subsequent patents that cite the focal patent, but not its prior art; $N_j$ is the number of subsequent patents that cite both the focal patent and its prior art; and $N_k$ is the number of subsequent patents that cite the focal patent's prior art but not the focal patent itself.

**Figure 1: Elements of the disruptiveness measure.**

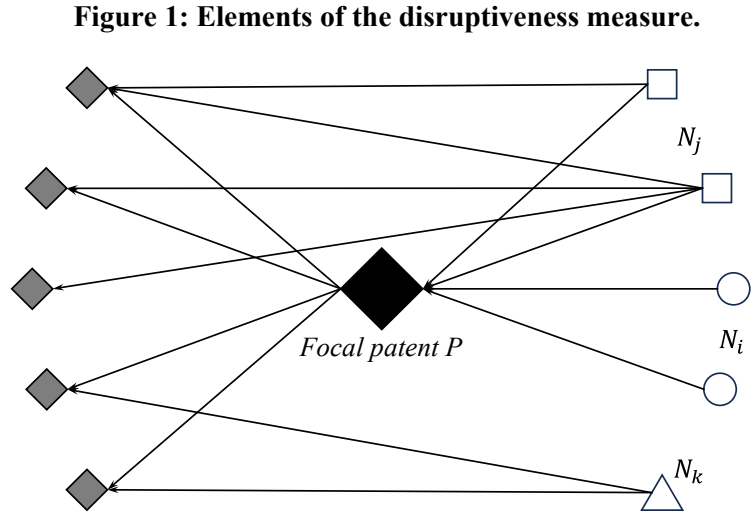

Source: The authors' drawing on the basis of Funk and Owen-Smith (2017).

Using this notation, the CD index proposed by Funk and Owen-Smith (2017) is given by

$$D_{F\_OS} = \frac{N_i - N_j}{N_i + N_j + N_k}. \tag{1}$$

Using (1), we can verify that the patents citing the focal patent but not its prior art ($N_i$) increase the disruptiveness of the focal patent, whereas the subsequent patents citing both the focal patent and its prior arts ($N_j$) diminish its disruptiveness. The number of those subsequent patents that build on the same prior art of the focal patent but that do not cite the focal patent ($N_k$) decreases the focal patent's disruptiveness. If forward citations to a patent do not at all acknowledge its prior art, the patent is disrupting its field and the focal patent is maximally disruptive ($D_{F\_OS} = 1$). In contrast, the focal patent is fully consolidating if $D_{F\_OS} = -1$, since such a patent strengthens the ties of subsequent patents to the focal patent's prior art.

Subsequent literature has suggested several modifications to the CD index. For example, Bornmann et al. (2020) modify the CD index by excluding the term $N_k$ from its denominator, since it seems to empirically dominate CD index values. Their suggested measure is

$$D_{BDTC} = \frac{N_i - N_j}{N_i + N_j}. \tag{2}$$

However, excluding $N_k$ is not uncontroversial since a focal patent may itself cite very impactful patents ($N_k$ adjusts for this possibility). Chen et al. (2021) and Leydesdorff and Bornmann (2021) argue that the potentially multiple roles of an invention should be acknowledged. They propose using two different metrics, one measuring disruptiveness and another consolidation:

$$D_{CSF-LTB} = \frac{N_i}{N_i + N_j + N_k} \tag{3a}$$

$$C_{CSF-LTB} = \frac{N_j}{N_i + N_j + N_k} \tag{3b}$$

Chen et al. (2021) discuss how and why a given invention can be both disruptive and consolidating at the same time and suggest measures for capturing this kind of multidimensionality. Bu et al. (2021) suggest an independence measure that is closely related to the CD index and to the other disruptiveness measures. It is given by

$$D_{BWH} = \frac{N_i}{N_i + N_j}. \tag{4}$$

This measure obtains higher values when a patent is not cited together with its prior art, i.e., when the focal patent is based on an invention that has been developed relatively independently from the earlier ones. We refer the reader to Leibel and Bornmann (2023) for further discussion of these and other related measures.

**Two dimensional and multidimensional measures:** Several scholars have noted that a discovery's capacity to impact and disrupt is likely to be multidimensional (Bu et al., 2021; Funk & Owen-Smith, 2017; Leibel & Bornmann, 2023; Wei et al., 2023). The citation-weighted mCD index of Funk and Owen-Smith (2017) aims at measuring both the impact and disruptiveness of a discovery. It is obtained by multiplying a discovery's CD index by its citation count and is given by

$$D^m_{F\_OS} = \frac{(N_i + N_j)(N_i - N_j)}{N_i + N_j + N_k} = \frac{(N_i^2 - N_j^2)}{N_i + N_j + N_k}, \tag{5}$$

showing that $D^m_{F\_OS} = F \times D_{F\_OS}$ where $F \equiv N_i + N_j$. Unfortunately, this measure is sensitive to variations in citation counts due to different scales of the CD index and the citation count (Leibel & Bornmann, 2023; Wei et al., 2023). This leads to counterintuitive outcomes since multiplying a discovery's CD-index by its citation count amplifies minor differences in the former. Specifically, we show that highly cited discoveries obtaining CD index values relatively close to zero may end up appearing extremely disruptive (or consolidating) in light of the mCD index.

Wei et al. (2023) propose keeping the CD index and citation counts separate and using cutoff values to determine, which discoveries are consistent with revolutionary science. They suggest zero as a cutoff for disruptiveness (the CD index) and $e^2 \approx 7.4$ as a cutoff for impact (the number of citations). A challenge with this approach is, however, that the appropriate thresholds are difficult to determine, leaving it open whether all those patents for which the CD index is greater than zero and for which the number of citations is more than eight should be regarded revolutionary science. For example, the thresholds do not necessarily allow identifying discoveries that are unique in these dimensions relative to other discoveries.

Bu et al. (2021) acknowledge that the impact of a research contribution (e.g., a publication or a patent) can be multidimensional. They argue that to characterize this, one must pay attention not only to the size of impact (e.g., the number of citations) but also to the qualitative nature of citation patterns, such as depth and breadth as well as dependence and independence of the citation impact. However, they propose neither a way to integrate these measures nor to link them to paradigm shifting discoveries.

Both Kelly et al. (2021) and Park et al. (2023) introduce measures for assessing patent quality by integrating novelty and impact into a single metric. Kelly et al. (2021) compare textual similarities between patents to discern a patent's novelty and impact. Their dynamic weighting methodology normalizes a term's importance over time, and they deploy an index to pinpoint "breakthrough" innovations. They employ citation data for validation and show that their text-based approach offers a consistent representation of inventions prior to 1945, when citation practices were less standardized.

The analysis of Park et al. (2023) allows for distinct meanings of backward and forward citation similarity, the incorporation of multiple words/topics in the text analysis (suggesting semantic multidimensionality), and the possibility of measuring quality over varying time horizons. To be more specific, Park et al. (2023) introduce two normalized versions of the CD index to measure disruption, adjusting for the number of citations by the focal paper/patent and accounting for field differences and time trends. They further explore an alternative disruption measure, $DI^*$, which makes the index less sensitive to small changes in the forward citation patterns of papers/patents that have no backward citations. They also use randomized citation networks to gauge disruption trends. Nonetheless, their empirical analysis primarily emphasizes the unidimensional measure of disruptiveness, the CD index in particular.

In contrast to these recent advances, we propose combining the existing disruption measures with the standard measures of impact and novelty, and using all three as inputs to an integrative measure that is easy to interpret. We turn to the development of such a measure now.

**3.2 A new multidimensional measure of paradigm shifting discoveries ($D_{FLEX}$)**

Our starting point for developing a multidimensional measure for paradigm shifting discoveries is the well-known mathematical formulae for the weighted generalized (power) mean i.e., $M_\sigma(z_1, \ldots, z_S) \equiv \left(\sum_{s=1}^{S} \alpha_s (z_s)^\sigma\right)^{\frac{1}{\sigma}}$, where $\sigma$ is a non-zero real number, $\alpha_s$ are non-negative weights that sum to one, and $z_s$ denotes a generic positive real number (see, e.g., Bullen, 2003). Our approach is inspired by Perry and Reny (2016) and Haley (2019). Perry and Reny (2016) present a new method for summarizing a researcher's publication record using the generalized mean. Haley (2019) uses a weighted power mean version of the Euclidean index for the same purpose. As Perry and Reny (2016) note, the weighted generalized mean is linked to the $l_\sigma$ norm in mathematics, the constant elasticity of substitution (CES) production function in economics, the poverty index (Foster et al., 1984), and the measure of income inequality (Atkinson, 1970). Ahmadpoor and Jones (2019) use a version of it to model how research teams generate patents and publications.

We use the generalized mean, $M_\sigma(z_1, \ldots, z_S)$, to capture the idea that the tendency of a discovery to be paradigm shifting depends on its (i) impact on future technologies, (ii) novelty, and (iii) ability to break with the past. Specifically, we use the percentages of the cumulative distribution of discoveries' impact, novelty, and tendency to break with the past as the normalized inputs ($z_s$) over which the generalized mean is calculated. We now turn to the measurement of the three dimensions, illustrate the importance of the normalization, and discuss the properties of $D_{FLEX}$ which motivate its use. In the discussion that follows, we let $x_s$ denote a raw measure and $G_s(x_s)$ the associated percentages of the cumulative distribution of $x_s$.

**Dimensions of $D_{FLEX}$ and their normalization:** The first measure captures discoveries' impact. For that, we use patents' citation counts ($x_1 \equiv F$), measured five years after they were granted. Second, to capture novelty, we use new keyword combinations (Gerken & Moehrle, 2012; Kim et al., 2016; Yoon & Park, 2005). We resort on the text-based approach, as it directly reflects a patent's written content (i.e. what the underlying idea is about) and as it is unclear which level of patent classes would properly capture novelty in our application (Righi & Simcoe, 2019). In our empirical analysis, we use the results from a text-based analysis of new keywords by Arts et al. (2021) – the number of new keyword combinations ($x_2 \equiv K$). Third, to capture a discovery's tendency to render prior work obsolete, we use the CD index ($x_3 \equiv D_{F\_OS}$) and, in alternative analyses, its variants ($D_{BDTC}$, $D_{CSF-LTB}$, and $D_{BWH}$). In line with prior work (Bornmann et al., 2020; Leydesdorff et al., 2021; Park et al., 2023) on the normalization of citation-based measures, we standardize the dimensions – $x_1, x_2, x_3$ –, over which the generalized mean is calculated by using the percentages of their respective cumulative distributions (Bornmann & Williams, 2020), i.e., using $G_s(x_s)$. This normalization means that the variables over which the generalized mean is calculated are $f_s = G_s(x_s)$ for $s = 1,2,3$ and that the raw variables $x_s$ are transformed to the same scale from zero to one and have the same unit of measurement. Crucially, this normalization also allows *ranking* patents relative to their comparators in terms of their impact, novelty, and tendency to break with the past and then to combine these rankings (using a comparable scale). For example, consider a focal patent with $x_1 = 6$ forward citations. Using the percentage of the cumulative forward citation distribution of the relevant pool of comparator patents gives $G_1(6)$, which is equal to the percentage of patents receiving $x_1 = 6$ or fewer forward citations than the focal patent. The larger the percentage of patents receiving fewer forward citations than the focal patent, the more impactful the focal patent. Similar reasoning applies to the normalization of the other dimensions ($x_2, x_3$).

How the cumulative distribution is calculated depends on the research question. If the goal is to learn about the development of paradigm shifting discoveries over time in different fields, the percentages of the cumulative distributions need to be calculated by field. This normalization enables an analysis of time patterns within each field.[7] It also is consistent with the goal of being able to quantify paradigm shifts within a broadly defined field. On the other hand, if the goal is to analyze whether there are more paradigm shifting discoveries in certain field at a given point in time, a normalization by (grant) years or cohorts is needed.

---

[7] Both Kelly et al. (2021) and Park et al. (2023) emphasize the need to adjust for evolving norms when analyzing patents over time (see also Kuhn et al., 2020).

**Definition of $D_{FLEX}$:** Using the notation for the generalized mean from above, the proposed measure is given by $M_\sigma(z_1, z_2, z_3) = M_\sigma(G_1(x_1), G_2(x_2), G_3(x_3))$. It can be written explicitly as follows:

$$D_{FLEX} \equiv \sqrt[\sigma]{\alpha_1(G_1(x_1))^\sigma + \alpha_2(G_2(x_2))^\sigma + \alpha_3(G_3(x_3))^\sigma}, \tag{6}$$

where $\sigma$ is a parameter of the generalized mean determining its shape (see below), $(\alpha_1, \alpha_2, \alpha_3)$ are non-negative weight parameters that sum to one, and $G_s(x_s)$ refers to the percentages of the cumulative distribution of $x_j$. Since each $G_s(x_s)$ is measured on the unit interval, $D_{FLEX}$ ranges from zero to one. It is increasing in the propensity of a discovery to be paradigm shifting, with $D_{FLEX} = 1$ indicating an extremely unique discovery in terms of *all* the three dimensions. As equation (6) shows, $D_{FLEX}$ is modular, in the sense that it combines distinct dimensions (impact, novelty, and disruptiveness) into a single, coherent measure. This property of $D_{FLEX}$ aligns with the concept of modularity used to understand the architecture of complex systems, as discussed by Brusoni et al. (2023). Just as modularity allows for the decomposition of complex systems into simpler, more manageable components, $D_{FLEX}$ captures the essence of paradigm shifting discoveries by integrating their key dimensions in a transparent and flexible manner.

**Properties of $D_{FLEX}$:** The generalized mean nests, as special cases, the arithmetic mean ($\sigma = 1$), geometric mean ($\sigma \to 0$), and harmonic mean ($\sigma = -1$). The arithmetic mean corresponds to the case when the included features are substitutes with each other, i.e., the case when a decrease in (say) forward citations can be compensated by an increase in novelty. A key property of the generalized mean is that parameter $\sigma$ determines *how* the dimensions (i.e., $G_1(x_1), G_2(x_2)$ and $G_3(x_3)$) produce – in *combination* – the value of $D_{FLEX}$. Holding other things constant, the dimension with the highest $G_s(x_s)$ becomes more influential when $\sigma$ increases; in the limit, when $\sigma \to \infty$, $D_{FLEX} = \max\{G_1(x_1), G_2(x_2), G_3(x_3)\}$ implying that only the maximum of the three dimensions matter. When $\sigma$ decreases, the dimension with the lowest $G_s(x_s)$ becomes more important; in the limit, when $\sigma \to -\infty$, $D_{FLEX} = \min\{G_1(x_1), G_2(x_2), G_3(x_3)\}$, implying that only the minimum of the three dimensions matter. In this latter case, $G_s(x_s)$ are strict complements, which means that a discovery's potential to be paradigm shifting is determined by its *weakest* dimension. As we will see, this turns out to be important in our empirical application.

It is worth pointing out additional properties of $D_{FLEX}$. First, it embeds the unidimensional measures as special cases. For example, if the capacity of a discovery to be paradigm shifting is solely determined by it being extremely highly cited, $D_{FLEX}$ can capture this: when $\alpha_2 = \alpha_3 = 0$ and $\sigma = 1$, we have $D_{FLEX} = G_1(x_1)$, meaning that $D_{FLEX}$ ranks discoveries solely in terms of their citations. Similarly, if $\alpha_1 = \alpha_2 = 0$ and $\sigma = 1$, then $D_{FLEX} = G_3(x_3)$ ranks discoveries solely in terms of their disruptiveness. Second, if $\sigma \to 0$, the limit of the weighted generalized mean is $D_{FLEX} = G_1(x_1)^{\alpha_1} G_2(x_2)^{\alpha_2} G_3(x_3)^{\alpha_3}$ (i.e., the geometric mean). To illustrate the connection of this to the mCD index, we can envisage that novelty does not matter ($\alpha_2 = 0$) and that $\alpha_1 = \alpha_3 = 1$. With these parameter restrictions, we obtain a (scaled) version of the mCD index, i.e., $D_{FLEX} = G_1(x_1) G_1(x_3)$, which is the product of the normalized citation count and the normalized CD index. The normalization removes the problem emerging from the different scales, suggesting that this special case of $D_{FLEX}$ provides an alternative way to solve the issue raised also by Wei et al. (2023).

The main goal of the empirical analysis is that we calibrate the parameters of $D_{FLEX}$ and explore whether complementarity or substitutability is supported by the data. We then make use of this calibration later to predict discoveries in unseen USPTO data (not used for the calibration).

## 4. EMPIRICAL ANALYSIS

### 4.1. Data

**Data sources:** We use patents granted by the United States Patent and Trademark Office (USPTO), for which an NBER (National Bureau of Economic Research) industry codes are available (obtained from patentsview.org). The patents are matched to text-based keyword information from Arts et al. (2021). Since the text-based keyword data are sparse for 1980 and 1981 and since the NBER codes are not available to us from May 2015 onwards, our sample runs from 1982/01 to 2015/06, containing 4,710,610 utility patents. When identifying forward citations and components of the disruption measures (i.e., $N_i$, $N_j$ and $N_k$), we include the citations irrespective of who (e.g., the assignee/inventor vs. patent examiner) has added them. Furthermore, we also account for the citations emerging from subsequent design and other non-utility patents.

We also use data collected from the National Inventors Hall of Fame (NIHF). We use NIHF inventions as manifestations of paradigm shifting discoveries and use USPTO patents related to them to calibrate the parameters of $D_{FLEX}$ in our empirical analysis. NIHF inductees are chosen based on their discoveries that have created a tangible impact on society and significantly advanced the fields of science and technology, serving as a testament to human ingenuity and inspiration for future generations.[8] The goal of these criteria is to ensure that only the most transformative innovations are honored – to sort out inventions that are not mere improvements but are foundational shifts in scientific and technological capability. NIHF's criteria revolve around the impact of an invention on the welfare of the nation, advancement of science, and the benefits it brings to the public. The patents protecting NIHF-recognized innovations represent paradigm shifting discoveries because they have laid an acknowledged foundation for new discoveries or revolutionized existing ones. Specifically, the nominees must have made a discovery that, according to a panel of experts in science, technology, and engineering, is considered "*a groundbreaking or a significant advancement*" in the nominee's field, suggesting that the patents recognized by NIHF represent instances of paradigm shifting discoveries. An example of a NIHF technology covered by a USPTO patent is the transistor (US2524035), which underpins all modern electronics and computing. Similarly, the NIHF-recognized discovery for CRISPR gene-editing technology (US8697359) has fundamentally changed the landscape of genetic engineering, enabling precise and efficient modification of DNA in organisms. These discoveries redefined what was feasible at the time.

Finally, we use patents identified as significant in the prior research and data on the so-called evergreen patents on pharmaceutical inventions included in the Orange Book (Durvasula et al., 2023, see also Appendix D). We use these data to illustrate how the proposed measure works.

**Descriptive statistics:** Table 1 shows the descriptive statistics for the five-year forward citations ($F$), the text-based measure for new keyword combinations ($K$), and the constituents used to calculate various disruptiveness measures, i.e., backward citations ($B$), $N_i$, $N_j$, and $N_k$.[9] As can be seen from Table 1, the mean number of forward citations, $F$, is 3.38, but the median (p50) is just 1. Novelty, as captured by $K$, also has a skewed distribution. The mean of $K$ is 123, but the median is only 19.

---

[8] See criteria for selection at the National Inventors Hall of Fame: https://www.invent.org/inductees/nominate-inventor (last visited 8 November 2023).

**Table 1: Descriptive statistics.**

|  | Mean | St. dev. | p5 | p25 | p50 | p75 | p95 | Min. | Max. |
|---|---|---|---|---|---|---|---|---|---|
| $F$ | 3.380 | 7.818 | 0 | 0 | 1 | 4 | 12 | 0 | 1,042 |
| $K$ | 123.527 | 2,259.131 | 0 | 2 | 19 | 73 | 366 | 0 | 1,440,810 |
| $B$ | 12.169 | 34.693 | 0 | 3 | 5 | 11 | 28 | 0 | 5,802 |
| $N_i$ | 1.926 | 4.538 | 0 | 0 | 1 | 2 | 7 | 0 | 496 |
| $N_j$ | 1.453 | 5.919 | 0 | 0 | 0 | 1 | 6 | 0 | 1,032 |
| $N_k$ | 158.007 | 582.757 | 0 | 8 | 30 | 96 | 621 | 0 | 64,827 |

**4.2. Empirical properties of existing disruption measures**

Despite going through challenges with the existing disruption measures in this section, our intention is *not* to criticize them. Quite the contrary, and as we wrote above, we suggest using them as an input to the proposed measure.

**Property #1 – Moderate impact and novelty**: Table 2 reports impact and novelty statistics for USPTO patents obtaining high values in terms of four existing disruption measures, i.e., $D_{F\_OS}$, $D_{BDTC}$, $D_{DSF-LTB}$, and $D_{BWH}$. In Panel A of Table 2, we use the absolute value of 0.90 as a threshold for high disruptiveness. In Panel B, we use the 90th percentile (p90) of each measure's distribution to capture patents for which relative disruptiveness is high.

As can be seen from Table 2, two of the scores, $D_{BDTC}$ and $D_{BWH}$, obtain their maximum values (of one) for more than 1.4 million patents, implying that they have mass points at the extreme right tails of their distributions (in line with Holst et al., 2024; see also our discussion below). Moreover, most of the patents with high disruption scores have less than five forward citations. The median (mean) number of forward citations among disruptive-looking patents is low, ranging from 1 to 2. Specifically, of the patents for which the CD index ($D_{F\_OS}$) value is exactly one (indicating maximal disruptiveness), 53% have one forward citation and 23% have two forward citations. The median number of new keyword combinations ranges from 19 to 27, indicating moderate degrees of novelty (at best).

---

[9] We winsorize the extreme values in the 99.9999% in the upper tail, i.e., replace the higher values of 4 observations with the value at the 99.9999% percentile.

**Table 2: Characteristics of disruptive patents as identified by the existing measures.**

| Panel A | Disruptiveness score ≥ 0.90 (absolute threshold) | | | |
|---|---|---|---|---|
| | $D_{F\_OS}$ | $D_{BDTC}$ | $D_{DSF-LTB}$ | $D_{BWH}$ |
| Number of disruptive patents | 191,526 | 1,486,255 | 191,659 | 1,514,410 |
| Share of disruptive patents | 0.040 | 0.315 | 0.040 | 0.321 |
| Conditional descriptive statistics | | | | |
| Share of patents with < 5 $F$ | 0.845 | 0.850 | 0.844 | 0.834 |
| Mean $F$ | 3.231 | 3.047 | 3.249 | 3.289 |
| Median $F$ | 2 | 2 | 2 | 2 |
| Mean $K$ | 211.35 | 125.929 | 211.151 | 126.401 |
| Median $K$ | 29 | 20 | 29 | 20 |
| **Panel B** | Disruptiveness score ≥ 90$^{th}$ percentile (relative threshold) | | | |
| | $D_{F\_OS}$ | $D_{BDTC}$ | $D_{DSF-LTB}$ | $D_{BWH}$ |
| Number of disruptive patents | 486,062 | 1,480,377 | 471,128 | 1,480,337 |
| Share of disruptive patents | 0.103 | 0.314 | 0.100 | 0.314 |
| Conditional descriptive statistics | | | | |
| Share of patents with < 5 $F$ | 0.732 | 0.853 | 0.704 | 0.853 |
| Mean $F$ | 4.603 | 2.888 | 4.892 | 2.888 |
| Median $F$ | 2 | 2 | 3 | 2 |
| Mean $K$ | 167.670 | 125.863 | 171.193 | 125.863 |
| Median $K$ | 27 | 20 | 27 | 20 |

**Property #2 – Concentration at extreme values**: Inspired by the findings reported in Table 2 above, by a comment made by Bornmann et al. (2020), and by Holst et al. (2024), we display the full distribution of $D_{F\_OS}$ in Figure 2.[10] As can be seen, there is a substantial difference between the number of patents that obtain the maximum value at the right end of the distribution and *just slightly* below. Relative to Holst et al. (2024), a novelty here is that we find that there also is a concentration at the right tail of the other proposed variants of the unidimensional disruption measures (see the corresponding distributions for $D_{BDTC}$, $D_{CSF-LTB}$ and $D_{BWH}$ in Appendix A). The mass point on the right is thus not just a characteristic of the CD index. Such non-smooth right tail behavior of disruptiveness is counter-intuitive and not desirable, as it suggests a discontinuity in the quality of the most exceptional discoveries (when, in reality, there is none).

**Figure 2: Histogram of $D_{F\_OS}$ (the CD index)**

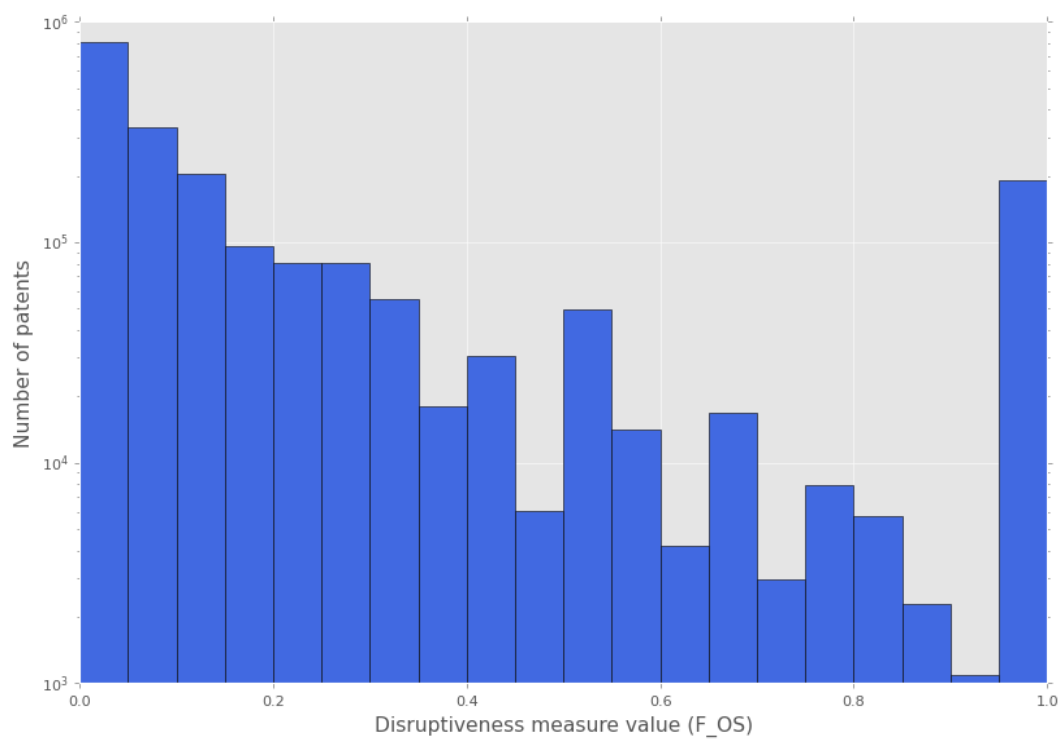

---

[10] To focus on the right tail of the distribution of these measures, we display their histograms conditional on the measures obtaining positive values.

**Property #3 – Different scales**: As we have noted earlier, the citation-weighted mCD index of Funk and Owen-Smith (2017) is sensitive to variation in citation counts due to different scales of the CD index and the citation count (see also Leibel & Bornmann, 2023). Figure 3 illustrates this. It displays patents with the CD index values in range [-0.20, 0.20] and their corresponding mCD-index values, with the dotted horizontal lines showing various percentiles of the mCD-index in the whole data. As can be seen, the scale difference amplifies minor differences in the CD index values for highly cited patents.

**Figure 3: Scatter plot of the mCD index and the CD index.**

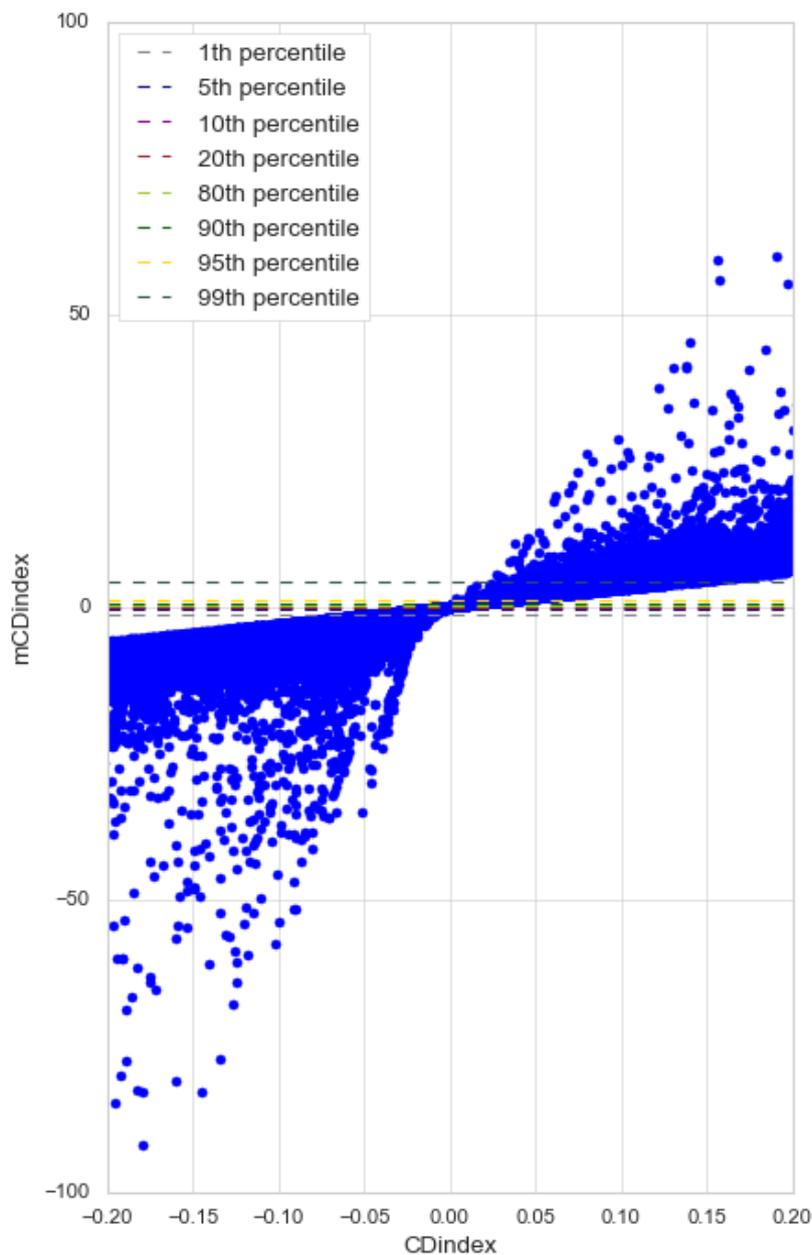

**Summary:** In sum, patents obtaining high values for $D_{F\_OS}$, $D_{BDTC}$, $D_{CSF-LTB}$, and $D_{BWH}$ are not, on average, highly cited and particularly novel. Furthermore, the distributions of these measures tend to concentrate at the extreme right end of the distributions. Finally, a citation-weighted version of the CD index (mCD) amplifies minor differences in the CD index. In line with the remarks made by, e.g., Leibel and Bornmann (2023) and Macher et al. (2023), we interpret these observations to speak for a need to find additional ways of identifying discoveries that are paradigm shifting. We now turn to such an endeavor and look for a parametrization of $D_{FLEX}$ applicable to the USPTO data.

**4.2. Estimation/calibration of the parameters of $D_{FLEX}$**

We use arguments from the prior literature and empirical analysis to find a plausible parameterization $(\alpha_1, \alpha_2, \alpha_3, \sigma)$ of $D_{FLEX}$.

**Theoretical arguments:** While the question of whether a discovery's impact, novelty and disruptiveness are complementary or substitutes has not been widely discussed, already the seminal writings, such as Kuhn (1962) and Dosi (1982), apparently alluded to the importance of complementarities. Echoing this, the work of Funk and Owen-Smith (2017) suggests – via the multiplication of the CD index by the number of forward citations – that a degree of complementarity is assumed. Several subsequent works follow this lead. In considering several alternative disruption indices and specifications, Bornmann et al. (2020) find that imposing a minimum threshold for the number of forward citations of a focal paper improves identification of disruptive contributions. This implies a complementary between forward citations and disruptiveness. Bu et al. (2021) suggest separating disruptiveness (or depth) and impact in a manner that implies complementary. While Wei et al. (2023) argue that the citation impact is an overly dominant term of the mCD index, their matrix for capturing revolutionary science, based on citation counts and disruptiveness scores, preserves the complementary assumption. Leahey et al. (2023) and Leibel and Bornmann (2023) also put forward arguments that speak for complementarity. While these considerations have no direct implications for the weight parameters of $D_{FLEX}$ (i.e., for $\alpha_1, \alpha_2, \alpha_3$), they clearly suggest that $\sigma$ ought to take on negative values. However, the literature does not seem to provide any guidance on the magnitude of the possible complementarity (and thus on suitable value of $\sigma$).

**Empirical analysis**: To proceed empirically, we must overcome the fundamental challenge of not having systematic data on paradigm shifting discoveries. To circumvent this problem, we use the USPTO patents connected to the NIHF discoveries. As explained earlier, they can be considered as manifestations of paradigm shifting discoveries. We can identify 105 patents protecting NIHF discoveries in our USPTO data. Among these patents, the patent grant years vary from 1982 to 2014, whereas the years of NIHF-induction (nomination) range from 1991 to 2023. Among these patents, the difference between the patent grant year and the induction year by NIHF is on average 19.8 years (median = 18). This means that the induction decision is based on a dispersed and rich set of qualitative and quantitative information that has cumulated over each discovery's (patent's) lifecycle and that has become available over time to those making the nomination. The NIHF nomination can be seen as a summary indicator of this information, allowing an identification of paradigm shifting discoveries.

We match these patents to a random subsample of USPTO patents (see Appendix B for the descriptive statistics). This means that we obtain a case-control sample with contaminated controls (Imbens & Lancaster, 1996, p. 146) – this is also called the case-population design. Unlike standard applications of choice-based sampling in econometrics (e.g., Hsieh et al., 1985; Imbens, 1992) and rare events analyses in statistics (e.g., King & Zeng, 2001), the case-population design consists of one subsample, selected fully on the outcome variable ("1") for which also the relevant covariates are observed, and another subsample drawn randomly from the whole population, for which *only* the covariates observed. Because there is no information on the outcome ("0" vs. "1") for these patents, the random subsample is "contaminated".

We use the percentages of the cumulative distributions of patents' impact ($x_1 = F$), novelty ($x_2 = K$) and disruptiveness ($x_3 = D_{F\_OS}$) as the input variables (i.e., $G_s(x_s)$ for $s = 1,2,3$). Empirically, the three dimensions are only moderately, if at all, correlated. For example, when calculated separately for each NBER industry, the pairwise correlation coefficients between the raw measures, $x_1$, $x_2$, and $x_3$, vary from -0.03 to 0.06 in the USPTO data. Therefore, they contain independent information about the potential of a discovery to be paradigm shifting. For our baseline empirical analysis, we calculate the percentages of the cumulative distributions of each dimension $x_s$ by each NBER industry. This choice allows us to focus on potential paradigm shifts that may transform specific industries.

To accommodate the specific nature of the case-control sample with contaminated controls, we use two econometric approaches. First, we estimate the parameters of $D_{FLEX}$ using a maximum likelihood (ML) estimator introduced by Steinberg and Cardell (1992).[11] Second, we use the generalized method of moments (GMM) approach of Imbens and Lancaster (1996, p. 150). These two approaches can accommodate any probability model, such as Logit, but we impose the structure of $D_{FLEX}$ from Equation (6).[12] The measure varies on the unit interval by design because of the normalization of the measures for citations, novelty, and disruptiveness. Finally, the estimators adjust by design for the fact that the model is fitted to a specific sample of patents (the case-control sample with contaminated controls). This property means that the estimated parameters are general in the sense that they apply to all USPTO patents (and not just the specific estimation sample). They can therefore be used to assign a likelihood of disruptive discovery to all patents based on their observable characteristics (i.e., based on each patent's $G_s(x_s)$ for $s = 1,2,3$).

---

[11] See also Imbens and Lancaster (1996), who use the ML estimator in their simulations.

**Estimation/calibration results**: Our results can be summarized succinctly (see Appendix B for details): First, we find systematic evidence that $\sigma$ tends to be negative, suggesting that the discoveries' citations, novelty, and disruptiveness are complementary. Second, both the ML and GMM estimations generate estimates of $\sigma$ that are large in absolute value. In particular, the estimates are negative enough to mean that, in practice, $D_{FLEX}$ can be approximated by the minimum over the three dimensions, i.e., by $D_{FLEX} \cong \min\{G_1(x_1), G_2(x_2), G_3(x_3)\}$, in our particular application (see Appendix B for details).[13] Quite remarkably, this empirical finding is nicely consistent with the arguments from the prior literature that speak for the importance of complementarities. Moreover, we have performed several sensitivity checks, such as used alternative numerical optimization methods, considered alternative random draws from the USPTO, and varied the inputs over which $D_{FLEX}$ is calculated, to make sure that $\sigma$ parameter indeed tends to negative values that are large enough in absolute terms to warrant the use of the minimum approximation (see Appendix B for details of these sensitivity checks).

---

[12] See Carvalho and Soares (2016) for an economic application, Rosenfeld (2017) for a political science application, and Lele (2009), Keating and Cherry (2004), and Rota et al. (2013) for applications in animal ecology.

[13] While in principle this limit is obtained only when $\sigma \to -\infty$, already much less negative values mean that the minimum is a good approximation. This also means that the values of the weights, i.e., $(\alpha_1, \alpha_2, \alpha_3)$, no longer bear on the values that $D_{FLEX}$ takes; see Appendix B for details and further discussion.

### 4.3. Properties of $D_{FLEX}$

Finding that the discoveries' citations, novelty, and disruptiveness are strictly complementary, i.e., $D_{FLEX} \cong min(G_1(x_1), G_2(x_2), G_3(x_3))$, is intuitively plausible. Specifically, it suggests that – whether a discovery is considered paradigm shifting – depends on its weakest dimension (relative to the other discoveries). To see this concretely, consider a hypothetical focal patent with $D_{FLEX}$ = *min*(0.40, 0.90, 0.95) = 0.40. The interpretation is that when compared to the other patents in the NBER industry class to which this patent belongs to, 40% have fewer forward citations than this focal patent; 90% of them have fewer new keyword combinations, and as many as 95% of them have a lower CD index value. When compared to its comparator patents, the capacity of this focal patent to be paradigm shifting is constrained by its low impact, despite it being quite unique in terms of its novelty and disruptiveness. The strict complementarity means that lack of impact cannot be compensated for by greater novelty or disruptiveness. This discussion demonstrates the gain of putting the three measures together as opposed to using them in isolation; as we will see in our empirical application, it is the weakest of the three dimensions that determines the tendency of a discovery to be paradigm shifting.

Figure 4 shows the complete histogram of $D_{FLEX}$, using USPTO data from 1982 to 2015.[14] Our interest is on how $D_{FLEX}$ works near the right tail of the histogram, which identifies the patents that are unique in terms of their impact, novelty, and tendency to break with the past. As can be seen, the distribution is well-behaved in its right tail (i.e., there is no "discontinuity" among the most exceptional technologies).[15] Specifically, there is no concentration of observations at the right end of the distribution, suggesting that the behavior of $D_{FLEX}$ is smooth *where it matters the most*.

---

[14] Among all the USPTO patents in our sample, the mean value of $D_{FLEX}$ is 0.26, whereas the median and the 90th percentile are, respectively, 0.17 and 0.60.

[15] We continue to use the percentages of the cumulative distributions of each dimension, calculated by each NBER industry. We show the distribution of $D_{FLEX}$ that use two alternative normalizations (year and year-industry) in Appendix C.

**Figure 4: Histogram of $D_{FLEX}$.**

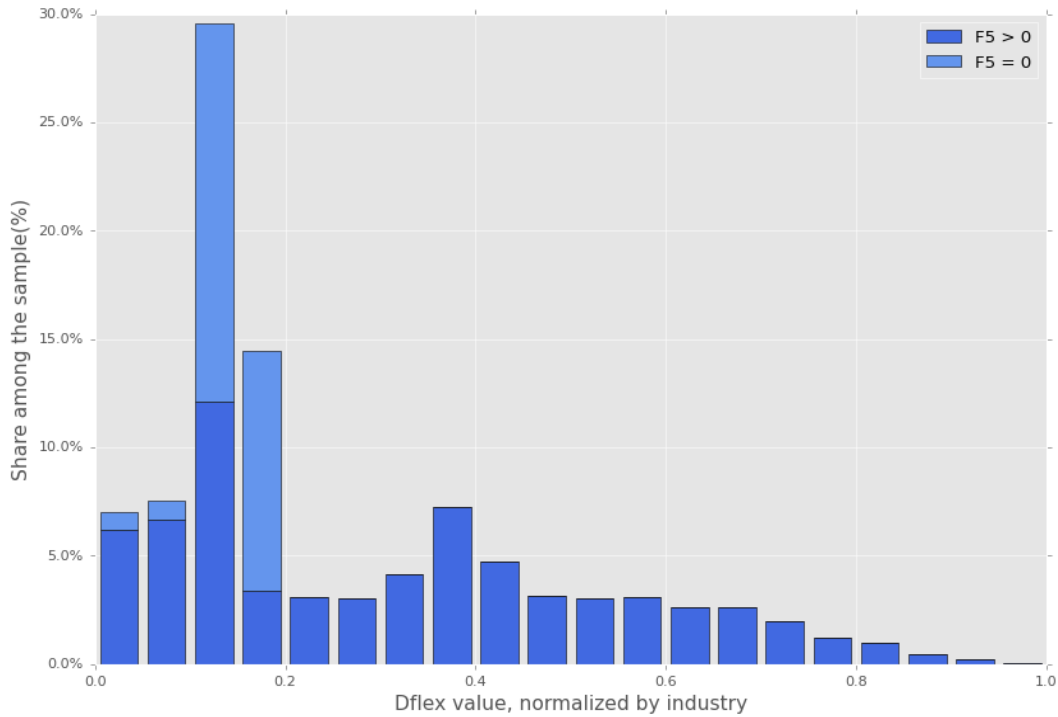

Table 4 displays the characteristics of patents that can be considered to be potentially paradigm shifting in light of their high $D_{FLEX}$ values; we use the same two thresholds as those in Table 2, i.e., the absolute value of 0.90 and the distribution's 90[th] percentile, p90. As can be seen, the absolute and the relative threshold values allow identifying a subset of patents that are clearly more impactful and novel and that have a greater tendency to render earlier technologies obsolete. For example, the average values of 0.95, 0.95 and 0.94 for $G_1(x_1)$, $G_2(x_2)$ and $G_3(x_3)$, respectively, mean that the patents with $D_{FLEX} \geq$ 0.90 belong, on average, to the highest 94[th] percentile of their respective forward citation, keyword, and CD index distributions. There are 10,804 USPTO patents with $D_{FLEX} \geq$ 0.90, representing 0.2% of all USPTO patents in our sample. This is an encouraging result, suggesting that the measure is capable of capturing patents that are *simultaneously* unique in all these dimensions. Using the relative threshold leads to a more relaxed criterion, but nevertheless identifies discoveries that are, on average, clearly impactful, novel, and disruptive, as can be seen from the respective means of $G_1(x_1)$, $G_2(x_2)$ and $G_3(x_3)$.

**Table 4: Characteristics of patents with high vs. low $D_{FLEX}$ values.**

|  | Absolute threshold (value) | | Relative threshold (percentile) | |
| --- | --- | --- | --- | --- |
|  | $D_{FLEX} < 0.90$ | $D_{FLEX} \geq 0.90$ | $D_{FLEX} <$ p90 | $D_{FLEX} \geq$ p90 |
| Number of patents | 4,699,806 | 10,804 | 4,932,270 | 471,340 |
| Share of patents | 0.997 | 0.002 | 0.899 | 0.100 |
| Conditional descriptive statistics | | | | |
| Share of patents with $F < 5$ | 0.791 | 0 | 0.826 | 0.461 |
| Mean of feature $x_1$ (= $F$) | 3.345 | 18.119 | 2.951 | 7.234 |
| Mean of feature $x_2$ (= $K$) | 121.144 | 1159.873 | 103.109 | 307.164 |
| Mean of feature $x_3$ (= $D_{F\_OS}$) | 0.071 | 0.545 | 0.055 | 0.226 |
| Mean of $G_1(x_1)$ | 0.498 | 0.949 | 0.466 | 0.797 |
| Mean of $G_2(x_2)$ | 0.498 | 0.954 | 0.465 | 0.811 |
| Mean of $G_3(x_3)$ | 0.498 | 0.941 | 0.464 | 0.816 |

### 4.4. Validation and case studies

**Long-term citations:** Paradigm shifting patents alter trajectories of subsequent inquiry. Because they constitute a break with the past and lay ground for further discoveries, their impact can come with a considerable lag. Building on this insight, Funk and Owen-Smith (2017) used long-term citations to identify truly impactful discoveries and then connected those to their disruption measures. Following their lead, we look at patents' impact well beyond the time window used to determine their $D_{FLEX}$ values.

To obtain Figures 5 and 6, we first group patents granted from 1982 to 1990 (558,172 patents) to ten bins of equal length by their $D_{FLEX}$ values and then calculate by the bin the forward citations they receive by 2015 – long after the time window used to determine their $D_{FLEX}$ values. When calculating the number of these long-term citations, we subtract the number of five-year citations, to make sure that there is no mechanical connection between patents' $D_{FLEX}$ values and their long-term impact. As Figure 5 suggests, the average number of long-term forward citations is clearly highest for patents with the highest $D_{FLEX}$ values, with a distinct increase in the rightmost tail (i.e., between the last two bins). Figure 6 shows the most impactful patents, defined as those subsequently receiving at least 120 forward citations – corresponding to the upper 1% tail of the *long-term* citation distribution. As can be seen, also the share of the most impactful patents grows with $D_{FLEX}$. The jump between the last two bins is even more pronounced, as the share of the high-impact patents nearly doubles between the last two bins of $D_{FLEX}$. Taken together, these two figures indicate that the highest values of $D_{FLEX}$ allow us to sort out discoveries that become extremely impactful over time (see Appendix C for similar graphs, drawn based on the percentiles of $D_{FLEX}$), based on patent data information that is available five years after they have been granted.

**Figure 5: Average number of long-term citations by bins of $D_{FLEX}$.**

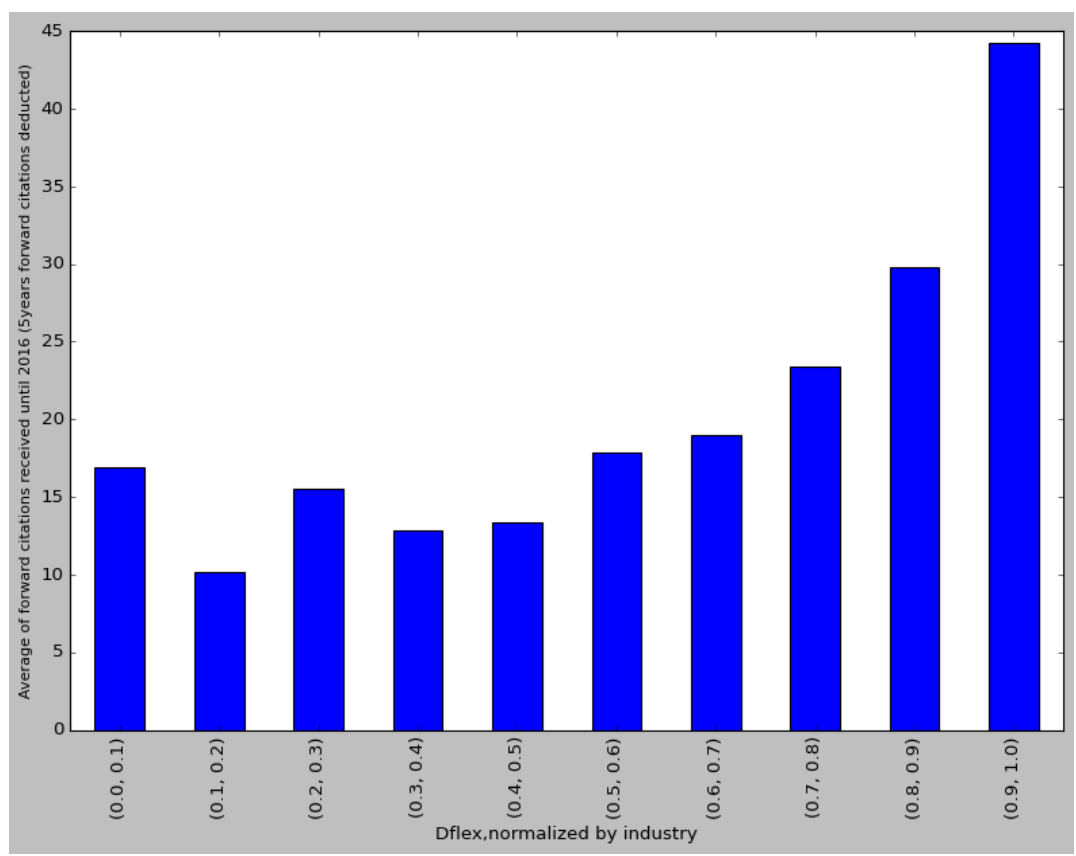

**Figure 6: Share of patents with long-term citations ≥ p99 by bins of $D_{FLEX}$.**

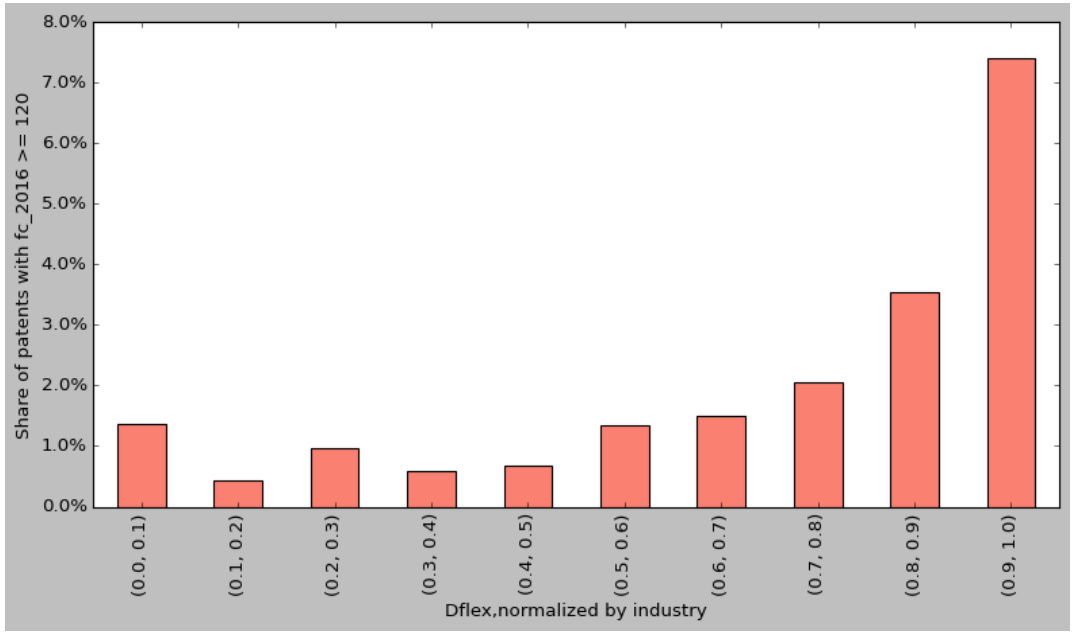

**Examples:** An alternative way to see how $D_{FLEX}$ performs is to consider known examples of patented discoveries that have been considered important elsewhere in the literature. In what follows, we build on Funk and Owen-Smith (2017), Chen et al. (2021), and Verhoeven et al. (2016) for this purpose. When evaluating how paradigm shifting the patents considered by these authors appear to be, we use the relative criterion (i.e., the 90$^{th}$ percentile value of $D_{FLEX} \Leftrightarrow D_{FLEX} \geq 0.60$) as well as the much more stringent absolute criterion (i.e., $D_{FLEX} \geq 0.90 \Leftrightarrow$ the 99,998$^{th}$ percentile value of $D_{FLEX}$).

Funk and Owen-Smith (2017) discuss a few illustrative patents. In their view, the most disruptive of them is the patent on polymerase chain reaction (US4683202) – commonly considered to be one of the founding inventions of modern biotechnology (Arts & Veugelers, 2014; Verhoeven et al., 2016). The USPTO data reveals that for this patent we have $(G_1(36), G_2(308), G_3(1)) = (0.989, 0.806, 0.973)$, yielding $D_{FLEX} = 0.806$. This value is based on the data that was available five years after the patent had been granted. It indicates that, at the time, the discovery was truly unique in terms of its impact and disruptiveness but its capacity to be paradigm shifting was constrained by its relatively moderate novelty. Specifically, it is paradigm shifting if we consider the relative criterion ($D_{FLEX} \geq 0.60$) but not if we consider the stricter absolute criterion ($D_{FLEX} \geq 0.90$). Another patent considered by Funk and Owen-Smith (2017) pertains to the atomic force microscope (US4724318), which is one of the central inventions in the field of nanotechnology. Although it was a radical improvement over prior art, the patent was considered by Funk and Owen-Smith (2017) to be neither clearly consolidating nor disruptive. In our data, we observe $(G_1(24), G_2(322), G_3(0.027)) = (0.987, 0.975, 0.638)$, giving $D_{FLEX} = 0.638$. This value suggests that the discovery is barely paradigm shifting according to the relative criterion (and clearly not according to the absolute criterion).[16]

Finally, Funk and Owen-Smith (2017) also point out two patents that intuitively seem disruptive but are not so according to their CD index. The first one is Eastman Kodak's patent for an early digital camera (US5016107), which contributed to a digital revolution in imaging (leading to Kodak's bankruptcy, as it missed the opportunity despite its early lead).[17] For this patent, we observe $(G_1(19), G_2(116), G_3(0.029)) = (0.980, 0.895, 0.646)$, resulting in $D_{FLEX} = 0.646$. This discovery seems to be paradigm shifting by a very narrow margin if the less stringent relative criterion is used. The second one, Stanford University's patent for node ranking in a linked database (US6285999), underlies Google's PageRank search algorithm and thus centrally contributed to modern ways of using the internet (and led Google to replace alternatives such as Altavista, Gopher, and Yahoo). For this patent, we observe $(G_1(31), G_2(194), G_3(0.162)) = (0.984, 0.931, 0.893)$, giving $D_{FLEX} = 0.893$. This patent is almost paradigm shifting also according to the stringer absolute criterion.

Chen et al. (2021) use MIT's patent for receiver-compatible enhanced television system (US5010405) as an example of a "destabilizing breakthrough" technology, while AsgrowSeed's patent for a hybrid soybean is their example of a "consolidating breakthrough" (US5576474). For the former, we find that $(G_1(9), G_2(294), G_3(0.474)) = (0.918, 0.971, 0.940)$, and $D_{FLEX} = 0.918$, suggesting that based on the data that was available five years after the patent grant, the discovery looked paradigm shifting both according to the relative and absolute criteria. For the consolidating one, our data reveals that $(G_1(18), G_2(0), G_3(-0.643)) = (0.960, 0.041, 0.003)$, and $D_{FLEX} = 0.003$.

---

[16] Funk and Owen-Smith (2017) also discuss a few examples that, in their consideration, are not disruptive but rather consolidating. These include a method applied in oil and gas drilling that enables better extraction of carbon monoxide and hydrogen from tar sands (US4573530) and a new genetically engineered soybean variety (US6958436). In our data, the values of $(G_1(x_1), G_2(x_2), G_3(x_3))$ are $(G_1(1), G_2(41), G_3(0.25)) = (0.391, 0.671, 0.892)$ and $(G_1(150), G_2(0), G_3(-0.847)) = (0.999, 0.041, 0.002)$, respectively, leading to $D_{FLEX} = 0.391$ for the former and $D_{FLEX} = 0.002$ for the latter. Thus, neither seems paradigm shifting.

[17] The explanation offered by Funk and Owen-Smith (2017) for an apparent lack of disruptiveness is that "… a single patent did not cover the advent of digital photography …" (p. 799).

Verhoeven et al. (2016) discuss both 3D printing and a "lab on a chip" as examples of patents that are both famous and reasonably novel. A key invention in the former was an apparatus for production of three-dimensional objects by stereolithography (US4575330). For this patent, we find that ($G_1$(22), $G_2$(186), $G_3$(0.241)) = (0.992, 0.947, 0.878), and that $D_{FLEX}$ = 0.878, suggesting that the discovery can be considered paradigm shifting if the relative criterion is used, but not quite if the stricter absolute criterion is used. The latter "lab on a chip" example refers to an invention in microfluidic manipulations (US5858195) used to separate, e.g., amino acids, peptides, and proteins in a way that made bio-chemical researchers 10–100 times faster on 100–10,000 times smaller samples. For this patent, we obtain ($G_1$(44), $G_2$(68), $G_3$(-0.030)) = (0.999, 0.566, 0.128), giving $D_{FLEX}$ = 0.13. Thus, despite its unusually high impact, the discovery seems not to be paradigm shifting.

**Case of pharmaceutical patents**: The Evergreen pharmaceutical patents dataset offers a distinction between "original" patents, protecting novel drugs, and "supplementary" patents, linked to novel drugs but aiming to extend initial intellectual property rights of original patents. The literature on evergreen patents (e.g., Feldman, 2018) departs from the notion that supplementary patents, *not* belonging to the initial intellectual property rights pool associated original patents, tend to be technical in nature and aim at extending the original patents' protection. This distinction suggests that the original patents ought to have greater impact, to be more novel and potentially also to be more disruptive – and, by extension, thus have more paradigm shifting features – than those used to extend the protection of a drug.

Starting with a dataset of 1,346 drugs included into the Orange Book (Durvasula et al., 2023) and their patents, we derive an estimation sample that consists of original and supplementary patents protecting the drugs (see Appendix D for details of the data). Using the resulting sample and a dependent variable indicating an original patent within a drug (= 1; a supplementary patent = 0), we can analyze whether $D_{FLEX}$ helps in distinguishing between the two types of drug patents. In a Logit regression, the coefficient for $D_{FLEX}$ is positive and highly statistically significant, implying a marginal effect of 0.15 (p-value < 0.01); we obtain similar results with a Linear Probability Model. While the difference between original and supplementary patents does not imply that the former are paradigm shifting, this additional empirical finding demonstrates the predictive and discriminatory power of $D_{FLEX}$.

**Summary:** Taken together, our analysis of long-term citations, examples of significant patents and pharmaceutical patents suggest that the proposed $D_{FLEX}$ measure works as intended. Specifically, it provides a transparent way of analyzing why some discoveries are, or are not, likely to be paradigm shifting.

### 4.5. Analyzing industry-specific trends using $D_{FLEX}$

Using several large datasets and $D_{F\_OS}$ (the CD index) as their main measure, Park et al. (2023) analyze how the disruptiveness research papers and patents has evolved over time. They focus on the temporal development of the average values $D_{F\_OS}$ in different industries, reporting that, over time, research papers and patents have become less disruptive, i.e., less likely to break with the past. Even though the recent analyses of Macher et al. (2024) and Holst et al. (2024) cast doubt on the validity of this finding, it is of interest to ask how the tendency of patents to be paradigm shifting has evolved over time by industry. To this end, we apply $D_{FLEX}$ to the percentages of each NBER industry's cumulative distributions, which allows us to explore, by industry, how $D_{FLEX}$ develops over time. Figure 6 displays our key results. Besides the average value of $D_{FLEX}$ (Panel A), we display the share of patents for which $D_{FLEX} > 0.90$ each year (Panel B), the value of 90$^{th}$ percentile of $D_{FLEX}$ by year (Panel C), and the share of patents for which $D_{FLEX} < 0.30$ (Panel D). The graphs display the annually calculated values of the measures for the full years from 1982 to 2014 and then for the first part of 2015.

Panels A–D in Figure 6 suggest the following: First, looking at Panel A, we find a steady downward trend in the annual means of $D_{FLEX}$ in each NBER industry. Panel B focuses on the right tail of $D_{FLEX}$ ($D_{FLEX} > 0.90$ in each year), which arguably better identifies paradigm shifting patents per industry, each year. This criterion implies that we only look at the patents whose impact, novelty and disruptiveness belong *at least* to their respective 90$^{th}$ percentiles, as calculated for each industry over period 1982/01-2015/06. We see a downward trend in the share of such patents. Panel C shows the value of the 90$^{th}$ percentile of $D_{FLEX}$, calculated separately for each year. It also shows that there is a downward trend in each NBER industry. Finally, Panel D shows that the share of patents with a low absolute value of $D_{FLEX}$ (< 0.3). The share of such patents has grown over time.

Because there is ongoing controversy in how backward citations and changes in patent laws should and could be accounted for (Holst et al., 2024; Macher et al., 2024), we refrain from making strong conclusions based on these findings. Taken at face value, they suggest that the tendency of patented discoveries to be paradigm shifting is in decline. In addition to measurement issues, these industry-specific trends may mirror technological regimes, which Corrocher et al. (2021) found to affect catching-up and leadership change in green technologies. Future research could explore the connections between technological regimes and the emergence of paradigm shifting discoveries across industries.

**Figure 6, Panel A: $D_{FLEX}$ over time, average.**

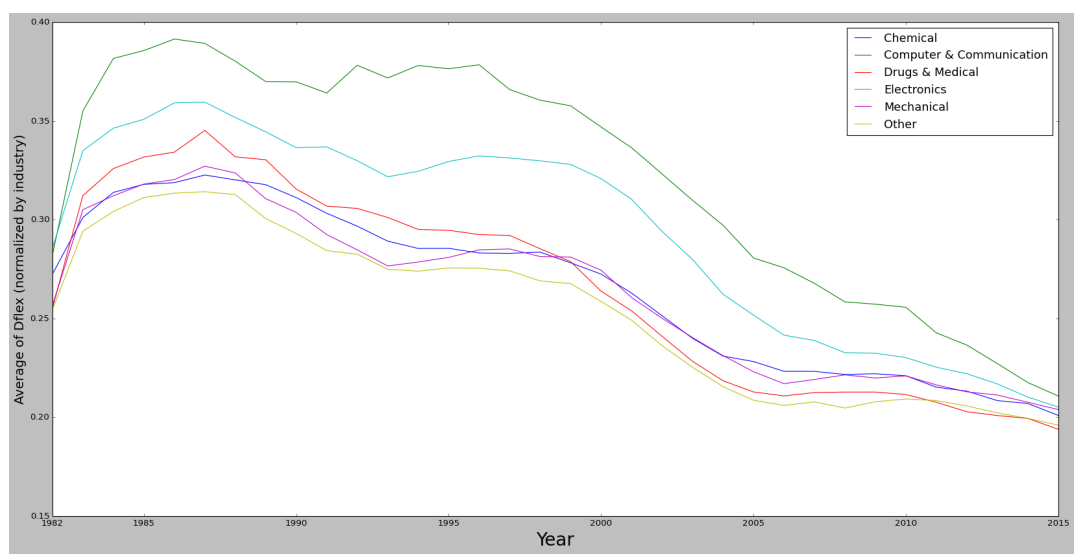

**Figure 6, Panel B:** $D_{FLEX}$ **over time, absolute value ≥ 0.90.**

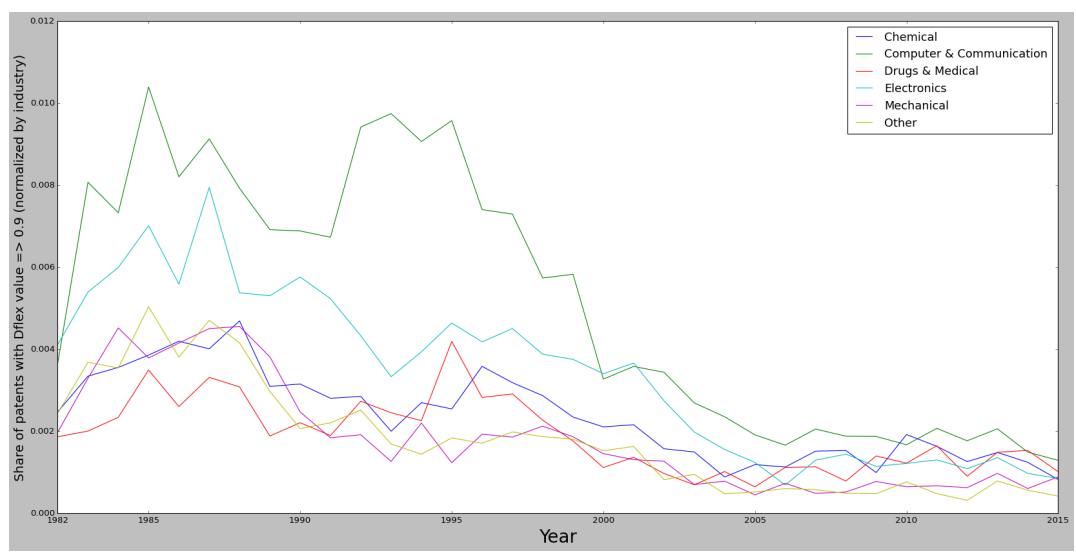

**Figure 6, Panel C:** $D_{FLEX}$ **over time, the value of the 90th percentile of** $D_{FLEX}$

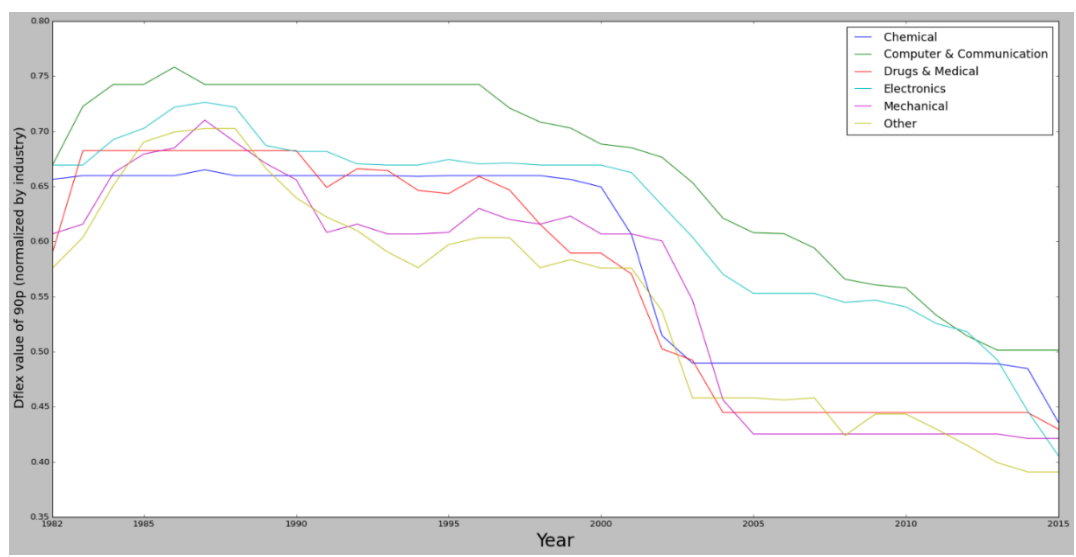

**Figure 6, Panel D:** $D_{FLEX}$ **over time, absolute value ≤ 0.30.**

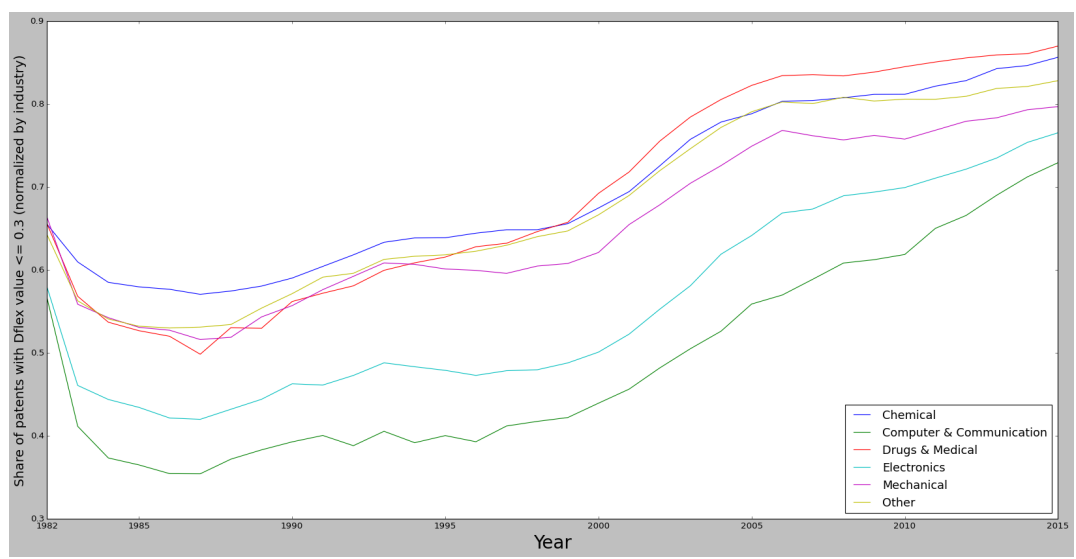

## 5. DISCUSSION

### 5.1 Limitations

A primary limitation of our empirical analysis is that we rely on patent data and use largely, but not solely, forward citations. To start with, using forward citations as inputs means that we cannot use $D_{FLEX}$ to identify paradigm shifting technologies in real time or ex ante. A buffer period of five years is required to accumulate such citations, causing a lag before $D_{FLEX}$, as used here, can be applied. However, for the purposes of understanding questions of whether there is more paradigm shifting going on in some industries than in others, or how paradigm shifting has developed over time, measures that make use of properties of forward citations are useful. Even for this purpose, there are limitations. For example, as Macher et al. (2024) point out, in the early years of the observation window, the calculation of the CD index may be affected by truncation bias that may distort its calculation. Moreover, patent data is unlikely to capture the entire spectrum of technological advance, given that not all significant inventions are patented. The variability in patenting practices across different regions and fields could subtly influence what $D_{FLEX}$ can quantify (our application solely uses USPTO data). This limits the generalizability of our findings.

Additionally, a notable limitation is our inability to broadly validate $D_{FLEX}$ because of the lack of data on paradigm shifting discoveries. Whereas we believe that our use of NIHF patents for the purpose of calibration is well motivated, the lack of validation data makes it difficult to thoroughly assess the accuracy of the proposed metric. Examining the relation between $D_{FLEX}$ and other indicators of revolutionary research could help in validating how well $D_{FLEX}$ works in different contexts. Such an analysis would also allow a better understanding of how the measure manages (or fails) to capture key theoretical constructs (Leibel & Bornmann, 2023; Wei et al., 2023).

Furthermore, it is not entirely clear which dimensions one should consider important for paradigm shifting discoveries. Whereas there are clear theoretical and empirical reasons to consider the discoveries' impact, novelty, and tendency to break with the past, this approach might miss other critical aspects, e.g., economic value. Such a need to have additional dimensions can be considered as an extension possibility, to which we turn next.

**5.2 Extension possibilities**

There are several ways one could extend $D_{FLEX}$ or tailor it for specific research questions. We can envisage, for example, the following:

**Additional dimensions**: A straightforward way to extend the measure is to add new dimensions, over which the generalized mean is calculated. Letting $S$ be the number of dimensions, we can write $D_{FLEX}$ as

$$D_{FLEX} = \sqrt[\sigma]{\sum_{s=1}^{S} \alpha_s G_s(x_s)^\sigma}, \qquad (7)$$

where $(\alpha_1, \ldots, \alpha_S, \sigma)$ are the parameters. This measure is flexible because the dimensions and their count, $S$, can depend on the context and the research question. Irrespectively of the original scale of $x_s$, using the percentage of the cumulative of distributions ensures that the unit of measurement of all included dimensions is the same. Moreover, using the percentage of the cumulative distribution allows considering their tail features. In this way, the measure could be extended to include new dimensions, such as depth and breadth of Bu et al. (2021), or non-technological aspects, such as discoveries' licensing revenues or market value.

**Complementary vs. substitutable dimensions**: Whereas we have argued that there are theoretical and empirical reasons to think that the impact, novelty, and breaking with the past are strongly complementary when paradigm shifting discoveries are considered, there can be research contexts in which the included dimensions are better viewed as substitutes. This suggests that $\sigma$ should take on a positive value. Specifically, increasingly larger positive values of σ give more emphasis to a discovery's most prominent dimension. Alternatively, the parameter can be calibrated or estimated using suitable context-specific data. Finding appropriate parameterizations of $D_{FLEX}$ for specific research purposes presents opportunities for future research.

**Weighting of the dimensions**: At times, there may be strong theoretical (a priori) reasons to weigh some of the dimensions, $x_s$, more than others. This can be achieved by varying the values of $(\alpha_1, \ldots, \alpha_S)$. When done transparently, this flexibility is a strength of the measure, allowing it to be tailored for the context.

### 5.3 Implications

Patents and scholarly publications contain information about whether and how science and technology advances. Being able to utilize that information matters both for managerial decisions and corporate strategy as well as for policy. For example, citation-based analyses of technologies can inform policy by enabling the identification of emerging key technologies (Huang et al., 2022), by helping to assess impacts of policies and regulation (Lin et al., 2021), by helping to evaluate the overall effectiveness of the patent system (Kwon, 2021), and by identifying and supporting active collaboration within the triple helix (Liu et al., 2023). Being able to measure the tendency of discoveries to be paradigm shifting provides insights on the evolution of industries and markets and is thus useful for firms and their management (see, e.g., Nelson & Winter, 1982; Sosa, 2011; Tushman & Anderson, 1986). Yet, lacking solid quantitative measures for paradigm shifting discoveries has been a persistent policy problem (Nagy et al., 2016), inhibiting both firms' and policymakers' ability to understand what is happening and to plan for the future.

Being able to identify potential paradigm shifting discoveries to inform policy questions requires that they can be identified and that their measurement can be adjusted to specific policy questions at hand. Transparent measurement builds trust and credibility in the assessment of scientific and technological advance and thereby enhances the ability of policymakers and stakeholders to make informed decisions. For example, to study what types of organizations, research milieus, and teams or which geographical locations are likely to produce paradigm shifting discoveries, one first needs a measure that can quantify the phenomenon of interest in a flexible way. Seen from this perspective, the proposed measure contributes to science of science and technology policy (Marburger et al., 2011; Powell et al., 2012).

## 6. CONCLUSIONS

Theory and empirics of economic growth show that longer-term improvements in human well-being are almost single-handedly driven by nurturing and applying new ideas. Whereas incrementally better new ideas and diffusion of old ideas are undoubtedly important, the role of paradigm shifting discoveries in the longer-term progress of humankind can hardly be overstated. Against this backdrop, it is indeed vital also to think how paradigm shifting technologies can be identified. In this study, we argue that paradigm shifting technologies (i) have a disproportionate impact on future technologies, (ii) are novel, and (iii) tend to induce a break in scientific and technological progress. Based on this insight, we incorporate these dimensions into a single, coherent measure and documented that these dimensions are strictly complementary.